\documentclass[english,notitlepage,showpacs,preprintnumbers,amsmath,amssymb,aps,twocolumn,nofootinbib]{revtex4-1}
\usepackage[T1]{fontenc}
\usepackage[latin9]{inputenc}
\usepackage{color}
\usepackage{array}
\usepackage{units}
\usepackage{multirow}
\usepackage{amsthm}
\usepackage{amsmath}
\usepackage{amssymb}

\makeatletter

\providecommand{\tabularnewline}{\\}

\@ifundefined{textcolor}{}
{%
 \definecolor{BLACK}{gray}{0}
 \definecolor{WHITE}{gray}{1}
 \definecolor{RED}{rgb}{1,0,0}
 \definecolor{GREEN}{rgb}{0,1,0}
 \definecolor{BLUE}{rgb}{0,0,1}
 \definecolor{CYAN}{cmyk}{1,0,0,0}
 \definecolor{MAGENTA}{cmyk}{0,1,0,0}
 \definecolor{YELLOW}{cmyk}{0,0,1,0}
}

\usepackage{microtype}
\usepackage{xspace}
\usepackage{amsfonts}
\usepackage{palatino}
\usepackage[]{algorithm2e}
\usepackage{MnSymbol}
\usepackage{bbm}
\usepackage{url}
\usepackage{ragged2e}
\edef\UrlBreaks{\do\-\UrlBreaks}
\usepackage{accents} 
\usepackage{microtype}

\usepackage{color}

\newcommand*{\mathcolor}{}
\def\mathcolor#1#{\mathcoloraux{#1}}
\newcommand*{\mathcoloraux}[3]{%
  \protect\leavevmode
  \begingroup
    \color#1{#2}#3%
  \endgroup
}

\newcommand{\hi}{$H_{\rm init}$\xspace}
\newcommand{\hf}{$H_{\rm final}$\xspace}

\makeatother

\usepackage{babel}
\begin{document}

\title{Crushing runtimes in adiabatic quantum computation with Energy Landscape
Manipulation (ELM): Application to Quantum Factoring}

\author{Richard Tanburn$^{1,\,}$}

\email{richard.tanburn@hertford.ox.ac.uk}

\affiliation{$^{1}$Mathematical Institute, Oxford University, OX2 6GG, Oxford,
UK. }

\author{Oliver Lunt$^{2,\,}$}

\email{oliver.lunt@trinity.ox.ac.uk}

\affiliation{$^{2}$Department of Physics, Oxford University, OX1 3PU, Oxford,
UK. }

\author{Nikesh S. Dattani$^{3,4,\,}$}

\email{dattani.nike@gmail.com}

\affiliation{$^{3}$School of Materials Science and Engineering, Nanyang Technological
University, 639798, Singapore,}

\affiliation{$^{4}$Fukui Institute for Fundamental Chemistry, 606-8103, Kyoto,
Japan}
\begin{abstract}
We introduce two methods for speeding up adiabatic quantum computations
by increasing the energy between the ground and first excited states.
Our methods are even more general. They can be used to shift a Hamiltonian's
density of states away from the ground state, so that fewer states
occupy the low-lying energies near the minimum, hence allowing for
faster adiabatic passages to find the ground state with less risk
of getting caught in an undesired low-lying excited state during the
passage. Even more generally, our methods can be used to transform
a discrete optimization problem into a new one whose unique minimum
still encodes the desired answer, but with the objective function's
values forming a different landscape. Aspects of the landscape such
as the objective function's range, or the values of certain coefficients,
or how many different inputs lead to a given output value, can be
decreased \emph{or} increased. One of the many examples for which
these methods are useful is in finding the ground state of a Hamiltonian
using NMR: If it is difficult to find a molecule such that the distances
between the spins match the interactions in the Hamiltonian, the interactions
in the Hamiltonian can be changed without changing the ground state
at all. We apply our methods to an AQC algorithm for integer factorization,
and the first method reduces the maximum runtime in our example by
up to 754\%, and the second method reduces the maximum runtime of
another example by up to 250\%. These two methods may also be combined.

\end{abstract}


\maketitle

\section{Introduction}

It has been proven that an adiabatic quantum computer (AQC) can simulate
any circuit-based quantum computation with only polynomial overhead
\cite{Mizel2007,Aharonov2008}, but so far it has been far easier
to build devices that implement adiabatic quantum algorithms than
circuit-based ones. For example, the most famous circuit-based quantum
algorithm (Shor's algorithm \cite{Shor1994,Shor1997}) has been celebrated
for more than 20 years, but never implemented with more than 7 qubits
\cite{DattaniBryans2014,Smolin2013,Vandersypen2001}. Contrarily,
quantum annealing devices have been used to find the ground states
of Hamiltonians with over 1000 qubits, and for many of the problems
solved on annealing devices such as in \cite{Ronnow2014,King2015},
no classical computer has ever found the solutions faster. Circuit-based
quantum algorithms have had nowhere near this level of success.

It is also strikingly easier to design an algorithm under the AQC
model of computation than the circuit-based model, for many problems.
For example, turning integer factorization into a Hamiltonian whose
ground state represents the prime factors can easily be understood
with only high-school education, or even less \cite{DattaniBryans2014,Burges2002}:
In fact, the whole algorithm can be explained in less than half a
page, including a fully worked example \cite{DattaniBryans2014}.

However, implementing an AQC algorithm is still not trivial:
\begin{enumerate}
\item The runtime increases significantly as the energy between the ground
state(s) and the lowest excited state(s) of the Hamiltonian becomes
small compared to the full range of the Hamiltonian's energies. 
\item If plenty of eigenstates of the Hamiltonian have energies close to
the ground state, the runtime for finding the ground state can be
much longer than if all excited states are at the highest energy levels. 
\item The required energy levels of the qubits, and the coupling strengths
between them, can be difficult to achieve in a real physical system. 
\end{enumerate}
We will show in this paper that it is possible to manipulate the size
of the energy gap between ground and excited states, or the number
of states at a given energy level, or the required coupling strengths
between the qubits. We refer to this as ELM (Energy Landscape Manipulation).

Various classical algorithms for solving discrete optimization problems
will also find a given problem easier or harder to solve depending
on the energy landscape of the Hamiltonian, except one may want to
call the Hamiltonian that needs to be minimized the ``objective function''
and the energy landscape as the ``values that the objective function
attains''. For example, the simulated annealing algorithm of \cite{Isakov2015}
performs better when the \emph{range }of the energy landscape is increased
(contrary to AQC, where it is usually preferred for the energy landscape
to have a small range). ELM can therefore also benefit classical discrete
optimization algorithms. 

Simply put, ELM can provide a new Hamiltonian (or objective function)
whose minimum still occurs at the same place, but is much easier to
find.

\section{The method}

\subsection{Terminology}

Suppose the solution to our problem is represented by the ground sate
of the Hamiltonian $H$. We use $|0\rangle,|1\rangle,\ldots,|n\rangle$
to denote the ground, 1$^{{\rm st}}$ excited, and $n^{{\rm th}}$
excited states of $H$ respectively, and their associated eigenvalues
(energies) are denoted: $E_{|0\rangle},E_{|1\rangle},\ldots,E_{|n\rangle}$
respectively. When there are multiple states with the same eigenvalue,
the number of states associated with eigenvalue $n$ is given by $N_{n}.$
We also define $E_{|{\rm max\rangle}}$ to be the largest eigenvalue,
and $E_{{\rm width}}\equiv E_{|{\rm max\rangle}}-E_{|0\rangle}$ to
be the \emph{spectral width} of the Hamiltonian (range of he objective
function). Likewise, $E_{{\rm gap}}\equiv E_{|1\rangle}-E_{|0\rangle}$
will be the \emph{spectral gap} of the Hamiltonian, and the \emph{spectral
ratio} will be $\mathfrak{\mathcal{R}}\equiv\nicefrac{E_{{\rm width}}^{2}}{E_{{\rm gap}}^{3}}$.
For time-dependent Hamiltonians, we will use the notation $E_{{\rm gap}}\equiv E_{{\rm gap}}(t)$.

\subsection{Motivation}

\subsubsection{Theoretical bound on the runtime of an adiabatic quantum computation}

In 1928, Born and Fock reported the adiabatic theorem, which is that
if a system is in a ground state of a Hamiltonian, and the Hamiltonian
is changed slowly enough, the system will stay in a ground state (of
the new Hamiltonian as it changes) \cite{Born1928}. But if the Hamiltonian
quickly changes into one in which one of its excited states has the
same energy as the ground state energy level of the original Hamiltonian,
the system might not have enough time to release the energy required
to stay in a ground state, and it will find itself in an excited state.
Therefore, one might wonder \emph{how slowly} one must change the
Hamiltonian, in order to ensure that the system is in the ground state
encoding the solution to the problem, after the Hamiltonian has at
last been fully transformed. 

If the Hamiltonian is transformed from $H_{{\rm init}}$ to $H_{{\rm final}}$
by the interpolant:

\begin{equation}
H(t)=\left(1-\frac{t}{T}\right)H_{{\rm init}}+\frac{t}{T}H_{{\rm final}},
\end{equation}
over a total ``runtime'' of time $T$, and we assume that the system
begins in $|\psi(0)\rangle$ a ground state of $H(0)\equiv H_{{\rm init}}$,
and we assume that the system evolves according to the non-relativistic
Schroedinger equation, the current best lower bound for the runtime
was derived in 2007 \cite{Jansen2007}. We present it in the notation
of \cite{Aharonov2008}: 
\begin{equation}
T\ge\max_{t}\frac{\left\Vert H_{{\rm final}}-H_{{\rm init}}\right\Vert ^{2}}{\epsilon E_{{\rm gap}}(t)^{3}},\label{eq:adiabatic theorem}
\end{equation}
which ensures that $|\psi(T)\rangle$, the solution to the Schroedinger
equation at time $T$, is close enough to the ground state $|0\rangle$
of $H_{{\rm final}}$ according to:

\begin{equation}
\sqrt{\langle\psi(T)|0\rangle}\le\epsilon.
\end{equation}
The matrix norm used is the \emph{spectral norm:}

\begin{equation}
\left\Vert H\right\Vert =\sqrt{{\rm max\, eigenvalue}\left(H^{\dagger}H\right)}.
\end{equation}

The starting Hamiltonian \hi is usually chosen to be one for which
the ground state is easy to realize in an experiment, so when attempting
to reduce the runtime, one generally has more freedom to manipulate
\hf than \hi. With this in mind, we look for ways to reduce $T$
by changing only the properties of \hf, while maintaining its defining
property: the encoding of the solution to the factorization problem
in its ground state.

One helpful fact is that in 2001, a bound for the largest eigenvalue
of $H_{{\rm final}}-H_{{\rm init}}$ was determined by the individual
eigenvalues of $H_{{\rm final}}$ and $H_{{\rm init}}$ \cite{Knutson2001}:

\begin{equation}
\left\Vert H_{{\rm final}}-H_{{\rm init}}\right\Vert \leq\lVert H_{{\rm final}}\rVert-\lVert H_{{\rm init}}\rVert.
\end{equation}
We can also set the zero of energy at the ground state of $H_{{\rm final}}$,
in which case $\lVert H_{{\rm final}}\rVert=E_{{\rm width}}$. We
also note that ${\rm min_{t}}\left(E_{{\rm gap}}(t)\right)\leq E_{{\rm gap}}(T)\equiv E_{{\rm gap}}$,
so Eq. \ref{eq:adiabatic theorem} leads to a less tight, but more
clear bound:

\begin{equation}
T\ge\frac{E_{{\rm width}}^{2}}{\epsilon E_{{\rm gap}}{}^{3}}=\mathcal{O}(\mathcal{R})
\end{equation}
Therefore we will make effort to reduce $E_{{\rm width}}$ and to
increase $E_{{\rm gap}}$, to ultimately reduce $\mathcal{O}(\mathcal{R})$
and hence reduce the lower bound on the runtime $T$.

\subsubsection{Properties that affect the runtime, but with no known bound (simulations
required)}

The bound on the runtime given in Eq. \ref{eq:adiabatic theorem}
is still not very tight. For example, two Hamiltonians may have the
same spectral ratio, but if one of them has $N_{1}=1$ (ie. only one
eigenstate in the first excited energy level $E_{|1\rangle}$) while
the other has $N_{1}=100$, intuition may expect that the runtime
would be longer in the latter case. Likewise, if low-lying excited
states (for example, $|2\rangle,\,|3\rangle,\,|4\rangle$) lie very
close to the ground state $|0\rangle$, one can imagine that the runtime
to find the ground state would be longer than if the distribution
of energies was such that these states were much further away. 

In the following sections, we present ways of changing the energy
landscape, with the hope of shifting the density of states away from
$E_{|0\rangle}$, and/or to increase the low-lying energy gaps $(E_{|n\rangle}-E_{|0\rangle}$
for small $n$) without increasing the spectral ratio $\mathcal{R}$
too much and while not changing the ground state(s) of the Hamiltonian.
Since it is not easy to analytically evaluate which changes to the
energy landscape will be more important in reducing the runtime, we
determine this by performing simulations.

\subsection{ELM using deductions: \emph{deduc-ELM}}

Our first method uses a \emph{deduction} to provide some control over
the energy landscape. For example, suppose it is known that the solution
to the problem requires $(x_{0},x_{1})=(0,0),(0,1)$ or $(1,1)$.
Such a deduction can be made by a local search, in which it is deduced
that setting $(x_{0},x_{1})=(1,0)$ will make it impossible to attain
the minimum, which is what it is that we are seeking. We have published
two papers this year on such deductions in discrete optimization problems
\cite{Tanburn2015a,DattaniBryans2014}. This deduction is equivalent
to the relation $x_{0}x_{1}=x_{0}$, since it holds under the valid
assignments and fails otherwise. With this relation, it is clear to
see that we can add 
\begin{equation}
\lambda(x_{0}x_{1}-x_{0})^{2}\label{eq:deducReducAdditionTerm}
\end{equation}
to our Hamiltonian for any $\lambda>0$, and the unique minimum will
not change. We know that the minimum we are looking for has either
$(x_{0},x_{1})=(0,0),(0,1)$ or $(1,1)$, in which case adding Eq.
\ref{eq:deducReducAdditionTerm} does not change the energy in any
way. However for invalid assignments of $(x_{0},x_{1})$, namely $(x_{0},x_{1})=(1,0)$,
adding Eq. \ref{eq:deducReducAdditionTerm} only adds more energy
to the Hamiltonian, which guarantees that such an assignment will
not have the lowest energy. In fact we will move all such invalid
states a distance of $\lambda$ further away from the global minimum,
affecting a quarter of all states in the search space.

In general, if we are able to deduce that $f=g$ for some polynomials
$f,g$ we can manipulate $H$ by adding $\lambda(f-g)^{2}$ for some
$\lambda>0$. This will move any state for which $f\neq g$ further
away from the global minimum. The effectiveness of such an addition
will be determined in turn by the energy landscape of $(f-g)^{2}$.
For more details on forming the addition term see our previous paper
which is entirely about using deductions (though not for ELM, but
for eliminating multi-qubit interactions in the problem Hamiltonian
without adding extra qubits) \cite{Tanburn2015b}.

\subsubsection{Application to the factoring problem}

For this section we pick $\lambda=1$ as the coefficient for all
of our deductions. This is in part for simplicity but mostly because
our aim is to move all of the states occupying $E_{|1\rangle}$ while
minimizing our impact on the spectral width. Any larger choice of
$\lambda$ would lead to a larger impact without pushing any more
terms away from $E_{|1\rangle}$.

We illustrate our method with the factorization of $841=29\times29$
which can be obtained by solving the following simultaneous equations:
\begin{eqnarray}
2p_{1}+p_{2}+q_{2} & = & 2z_{23}+4z_{24}\label{eq:deducELMEq1}\\
p_{1}p_{2}+p_{1}q_{2}+p_{3}+q_{3}+z_{23} & = & 2z_{34}+4z_{35}+1\\
p_{1}p_{3}+p_{1}q_{3}+p_{2}q_{2}+z_{24}+z_{34}+2 & = & 2z_{45}+4z_{46}\\
2p_{1}+p_{2}q_{3}+p_{3}q_{2}+z_{35}+z_{45} & = & 2z_{56}+4z_{57}\\
p_{2}+p_{3}q_{3}+q_{2}+z_{46}+z_{56} & = & 2z_{67}+4z_{68}+1\\
p_{3}+q_{3}+z_{57}+z_{67} & = & 2z_{68}+4z_{79}\label{eq:deducELMEqPen}\\
z_{68}+z_{79} & = & 1.
\end{eqnarray}
 It should be noted that the above equations are the result of elementary
simplification of the original factorization equations, as described
in \cite{DattaniBryans2014}. The Hamiltonian could also be simplified
much further, but then we would not have enough qubits to keep this
example interesting. See \cite{DattaniBryans2014,Xu2012} for details
on forming the original equations.

We then create our first Hamiltonian 
\begin{equation}
H_{0}=(2p_{1}+p_{2}+q_{2}-2z_{23}-4z_{24})^{2}+...+(z_{68}+z_{79}-1)^{2}
\end{equation}
 which encodes the single ground state (note that there will only
be one ground state since $841$ is a perfect square). A summary of
$H_{0}$'s energy landscape can be found in Table \ref{tab:factorizationTableDeduc}.

To perform ELM by deductions, we first need some deductions. Our first
set is obtained by looking at Eq. \ref{eq:deducELMEq1}. We notice
that if $z_{24}=1$ then we must have $p_{1}=p_{2}=q_{2}=1$. This
information can be encoded by
\begin{eqnarray}
z_{24}(1-p_{1}),\, & z_{24}(1-p_{2}),\mbox{\,\ and } & z_{24}(1-q_{2})
\end{eqnarray}
 respectively. We can then consider
\begin{eqnarray}
H_{1} & = & H_{0}+z_{24}(3-p_{1}-p_{2}-q_{2})
\end{eqnarray}
 and analyze our impact on the energy landscape (see Table \ref{tab:factorizationTableDeduc}).

Furthermore, we can then make an identical analysis of Eq. \ref{eq:deducELMEqPen}
to obtain our last Hamiltonian
\begin{equation}
H_{2}=H_{1}+z_{79}(4-p_{3}-q_{3}-z_{57}-z_{67}).
\end{equation}
With this we were able to manipulate the energy landscape of the problem
enough to have a huge impact on the spectral ratio $\mathcal{R}$,
as can be seen in Table \ref{tab:factorizationTableDeduc}.

\begin{center}
\begin{table}
\protect\caption{Energy landscapes for a 17-qubit factorization of $841$ for various
Hamiltonians $H$ with equivalent ground states.\label{tab:factorizationTableDeduc}}

\begin{centering}
\begin{tabular*}{1\columnwidth}{@{\extracolsep{\fill}}ccccccccccc}
\hline 
\noalign{\vskip2mm}
$H$ & $E_{{\rm gap}}=E_{|1\rangle}$ & $n_{1}$ & $E_{|2\rangle}$ & $n_{2}$ & $E_{|3\rangle}$ & $n_{3}$ & $E_{|4\rangle}$ & $n_{4}$ & $E_{|{\rm {max}\rangle}}$ & $\mathcal{R}$\tabularnewline[2mm]
\hline 
\noalign{\vskip2mm}
$H_{0}$ & $1$ & $4$ & $2$ & $5$ & $3$ & $14$ & $4$ & $38$ & $166$ & $27\,556$\tabularnewline
$H_{1}$ & $1$ & $2$ & $2$ & $7$ & $3$ & $14$ & $4$ & $36$ & $169$ & $28\,561$\tabularnewline
$H_{2}$ & $2$ & $8$ & $3$ & $10$ & $4$ & $32$ & $5$ & $54$ & $171$ & $3\,655$\tabularnewline[2mm]
\hline 
\end{tabular*}
\par\end{centering}

\centering{}\rule[-0.5ex]{1\columnwidth}{0.5pt}
\end{table}

\par\end{center}

We can also see that our set of deductions from Eq. \ref{eq:deducELMEq1}
managed to push away two of the four states originally occupying energy
level $E_{|1\rangle}$, and even some states occupying other excited
energy levels were pushed away, with a very small impact to the spectral
width. Introducing the second set of deductions moves the remaining
two states so that $\mathcal{R}$ is almost halved from its initial
value, with minimal impact on $E_{|{\rm {max}\rangle}}$.

\subsection{ELM using multiplicity: \emph{multiplicity-ELM}}

Our second ELM method is useful for any problem that is built from
a system of equations of discrete-valued variables. Suppose we begin
with a solvable system of binary equations:
\begin{eqnarray}
x_{1}+x_{2} & = & x_{3}+1\label{eq:toyExampleEq1}\\
x_{1}+x_{1}x_{2} & = & 2x_{2}x_{3}+x_{2}+1.\label{eq:toyExampleEq2}
\end{eqnarray}
We know that we can form a Hamiltonian by moving all terms to the
left side, then squaring and summing \cite{DattaniBryans2014}: 
\begin{equation}
H_{0}=(x_{1}+x_{2}-x_{3}-1)^{2}+(x_{1}+x_{1}x_{2}-2x_{2}x_{3}-x_{2}-1)^{2}.
\end{equation}
$H_{0}$ will achieve its global minimum of $0$ at exactly the solutions
to the original equations. However, we do not need to add these equations
with equal weight. For any positive $\lambda_{1},\lambda_{2}$, an
equally valid Hamiltonian would be: 
\begin{equation}
H_{1}=\lambda_{1}(x_{1}+x_{2}-x_{3}-1)^{2}+\lambda_{2}(x_{1}+x_{1}x_{2}-2x_{2}x_{3}-x_{2}-1)^{2}.
\end{equation}

Our first strategy is to choose $\lambda_{1},\lambda_{2}$ such that
the maximum energy contributed by each equation is the same. For example
Eq. \ref{eq:toyExampleEq1} contributes $(x_{1}+x_{2}-x_{3}-1)^{2}$
to the Hamiltonian, and therefore adds a maximum energy of $(0+0-1-1)^{2}=4$,
and Eq. \ref{eq:toyExampleEq2} contributes a maximum energy of $(0+0-2-1-1)^{2}=16$.
To address this, we will set $\lambda_{1}=4\lambda_{2}$. In general
the motivation for this method is that some equations contain few
variables and small coefficients, so the largest energy contribution
over the entire energy landscape will be quite small. We can then
add more weight to these equations, hoping that we do not impact the
spectral width too severely, whilst pushing many (hopefully low-energy)
states further away from the ground state. 

Specifically, let $E_{1}=\max(x_{1}+x_{2}-x_{3}-1)^{2}=4$ be the
maximum error contributed by the first equation, and similarly let
$E_{2}=16$ be the maximum contribution of the second. Furthermore
let $E_{{\rm max}}=\max(E_{1},E_{2})$. We then choose $\lambda_{1}=E_{{\rm max}}/E_{1}=4$
and $\lambda_{2}=E_{{\rm max}}/E_{2}=1$. Then, by analyzing the
energy landscape we see a great improvement in spectral ratio in Table
\ref{tab:multELMToy} by using our method rather than the naive approach.

\begin{table}
\protect\caption{Energy landscape for the toy example.\label{tab:multELMToy}}

\begin{tabular*}{1\columnwidth}{@{\extracolsep{\fill}}ccccccc}
\hline 
\noalign{\vskip2mm}
Hamiltonian & $E_{{\rm gap}}=E_{|1\rangle}$ & $n_{1}$ & $E_{|2\rangle}$ & $n_{2}$ & $E_{|{\rm {max}\rangle}}$ & $\mathcal{R}$\tabularnewline[2mm]
\hline 
\noalign{\vskip2mm}
$H_{0}$ & $1$ & $2$ & $2$ & $1$ & $17$ & $289$\tabularnewline
$H_{1}$ & $4$ & $4$ & $5$ & $1$ & $20$ & $6.25$\tabularnewline[2mm]
\hline 
\end{tabular*}

\centering{}\rule[-0.5ex]{1\columnwidth}{0.5pt}
\end{table}

In general, finding the absolute maximum energy contribution of an
equation can be costly, since maximization of an arbitrary polynomial
over the integers is a hard problem \cite{Anthony2015}.  However
for the factorization problem, we are able to maximize them easily
as each variable only appears once in each equation. To maximize the
energy contribution of each equation, we first check that each side
is fully expanded and that all of the coefficients are positive (by
moving terms to the other side if necessary). We then maximize the
left hand side by setting all of the variables to $1$ and minimize
the right hand side by setting all of its variables to $0$. The difference
between the sides can now be treated as a candidate for the maximum.
The other candidate is obtained by setting the variables in the left
hand side to $0$ and the rest to $1$. Squaring of the larger of
these two candidate maxima gives us the maximum possible energy contribution
of the equation.

It should be noted that in general, the ratio $E_{{\rm max}}/E_{i}$
will not be an integer. For simplicity, in this paper we will only
work with integer coefficients, and will therefore round all such
ratios for all implementations of ELM. However, non-integer ratios
are equally valid.

\subsubsection{Application to integer factorization}

We exhibit the method on the factorization problem: $551=19\times29$.
By generating the factorization equations and performing some very
simple deductions to reduce the problem slightly (such as: $x_{1}+x_{2}=0$
implies that $x_{1}=x_{2}=0$) we obtain the following set of simultaneous
equations:
\begin{eqnarray}
p_{1}+q_{1} & = & 1\label{eq:factorEq1}\\
p_{2}+q_{2} & = & 1\label{eq:factorEq2}\\
p_{1}q_{2}+p_{2}q_{1}+p_{3}+q_{3} & = & 2z_{34}+4z_{35}\label{eq:factorEq3}\\
p_{1}q_{3}+p_{2}q_{2}+p_{3}q_{1}+z_{34}+2 & = & 2z_{45}+4z_{46}\label{eq:factorEq4}\\
p_{1}+p_{2}q_{3}+p_{3}q_{2}+q_{1}+z_{35}+z_{45} & = & 2z_{56}+4z_{57}+1\label{eq:factorEq5}\\
p_{2}+p_{3}q_{3}+q_{2}+z_{46}+z_{56} & = & 2z_{67}+4z_{68}\label{eq:factorEq6}\\
p_{3}+q_{3}+z_{57}+z_{67} & = & 2z_{78}+4z_{79}\label{eq:factorEq7}\\
z_{68}+z_{78}+1 & = & 2z_{89}\label{eq:factorEq8}\\
z_{79}+z_{89} & = & 1.\label{eq:factorEq9}
\end{eqnarray}
We then form the naive Hamiltonian 
\begin{eqnarray}
H_{0} & = & (p_{1}+q_{1}-1)^{2}+...+(z_{79}+z_{89}-1)^{2}.
\end{eqnarray}
A quick check confirms that Eqs. \ref{eq:factorEq1}-\ref{eq:factorEq9}
do not have any variables in both the left and right hand sides, even
after our judgments, so we can carry out the algorithm described above
for maximizing the energy contribution from each equation. We sum
the appropriate coefficients and then choose $\lambda_{i}$ using
the formula
\begin{equation}
\lambda_{i}=\left\lceil E_{{\rm max}}/E_{i}\right\rceil .\label{eq:approxLambdaFormula}
\end{equation}

\begin{center}
\begin{table}
\protect\caption{Maximum energy contributed by each equation, for the factorization
$551=19\times29$.}

\begin{centering}
\begin{tabular*}{1\columnwidth}{@{\extracolsep{\fill}}clc}
\hline 
\noalign{\vskip2mm}
\multirow{1}{*}{Equation ($i$)} & \multirow{1}{*}{Maximum Energy $E_{i}$ } & $\lambda_{i}=\left\lceil E_{{\rm max}}/E_{i}\right\rceil $\tabularnewline[2mm]
\hline 
\noalign{\vskip2mm}
\ref{eq:factorEq1} & $\max(2,1)^{2}=4$ & $13$\tabularnewline
\ref{eq:factorEq2} & $\max(2,1)^{2}=4$ & $13$\tabularnewline
\ref{eq:factorEq3} & $\max(4,6)^{2}=36$ & $2$\tabularnewline
\ref{eq:factorEq4} & $\max(6,6)^{2}=36$ & $2$\tabularnewline
\ref{eq:factorEq5} & $\max(6,7)^{2}=49$ & $1$\tabularnewline
\ref{eq:factorEq6} & $\max(5,6)^{2}=36$ & $2$\tabularnewline
\ref{eq:factorEq7} & $\max(4,6)^{2}=36$ & $2$\tabularnewline
\ref{eq:factorEq8} & $\max(3,2)^{2}=9$ & $6$\tabularnewline
\ref{eq:factorEq9} & $\max(2,1)^{2}=4$ & $13$\tabularnewline[2mm]
\hline 
\end{tabular*}.
\par\end{centering}

\centering{}\rule[-0.5ex]{1\columnwidth}{0.5pt}
\end{table}

\par\end{center}

\noindent This results in the ELM-adjusted Hamiltonian:

\begin{widetext}

{\small{}
\begin{eqnarray}
H_{1} & = & {\color{red}{\normalcolor \mathbf{{\color{red}13}}}}(p_{1}+q_{1}-1)^{2}+{\normalcolor \mathbf{{\color{red}13}}}(p_{2}+q_{2}+1)+{\normalcolor {\color{red}\mathbf{2}}}(p_{1}q_{2}+p_{2}q_{1}+p_{3}+q_{3}-2z_{34}-4z_{35})^{2}+{\normalcolor {\color{red}\mathbf{2}}}(p_{1}q_{2}+p_{2}q_{1}+p_{3}+q_{3}-2z_{34}-4z_{35})^{2}\nonumber \\
 &  & +{\normalcolor {\color{red}\mathbf{1}}}(p_{1}+p_{2}q_{3}+p_{3}q_{2}+q_{1}+z_{35}+z_{45}-2z_{56}-4z_{57}-1)^{2}+{\normalcolor {\color{red}\mathbf{2}}}(p_{2}+p_{3}q_{3}+q_{2}+z_{46}+z_{56}-2z_{67}-4z_{68})^{2}\\
 &  & +{\normalcolor {\color{red}\mathbf{2}}}(p_{3}+q_{3}+z_{57}+z_{67}-2z_{78}-4z_{79})^{2}+{\normalcolor {\color{red}\mathbf{6}}}(z_{68}+z_{78}+1-2z_{89})^{2}+{\normalcolor {\color{red}\mathbf{13}}}(z_{79}+z_{89}-1)^{2}.\nonumber 
\end{eqnarray}
}{\small \par}

\end{widetext}

Again,  we show the spectral gap, width, ratio and nature of the
first few excited states in Table \ref{tab:factorizationTable511}.

We achieved a large reduction in the number of local minima close
to the ground state. However, we have doubled the spectral gap and
slightly more than doubled the spectral width, resulting in a slight
increase in spectral ratio. This is undesirable for computation by
an adiabatic quantum computer, so we present a small refinement.

\subsubsection{Refinement\label{sub:factorizationMod}}

In the factoring problem, we know \cite{Burges2002,DattaniBryans2014}
that the Hamiltonians always have the property that $E_{|1\rangle}=1$,
so if we are able to shift $E_{|1\rangle}$ up by just $1$, while
minimizing impact to $E_{|{\rm max}\rangle}$ at all costs, we may
be more likely to reduce the spectral ratio. We present an alternative
choice of $\lambda_{i}$ that achieves such a goal:
\begin{eqnarray}
\lambda_{i} & = & \begin{cases}
1, & E_{i}=E_{{\rm max}}\\
2, & E_{i}\neq E_{{\rm max}}
\end{cases}.\label{eq:indicatorLambdaFormula}
\end{eqnarray}
We then define the Hamiltonian produced according to this scheme:
\begin{widetext}{\small{}
\begin{eqnarray}
H_{2} & = & {\normalcolor {\color{red}\mathbf{2}}}(p_{1}+q_{1}-1)^{2}+{\normalcolor {\color{red}\mathbf{2}}}(p_{2}+q_{2}+1)+{\normalcolor {\color{red}\mathbf{2}}}(p_{1}q_{2}+p_{2}q_{1}+p_{3}+q_{3}-2z_{34}-4z_{35})^{2}+{\normalcolor {\color{red}\mathbf{2}}}(p_{1}q_{2}+p_{2}q_{1}+p_{3}+q_{3}-2z_{34}-4z_{35})^{2}\nonumber \\
 &  & +{\normalcolor {\color{red}\mathbf{1}}}(p_{1}+p_{2}q_{3}+p_{3}q_{2}+q_{1}+z_{35}+z_{45}-2z_{56}-4z_{57}-1)^{2}+{\normalcolor {\color{red}\mathbf{2}}}(p_{2}+p_{3}q_{3}+q_{2}+z_{46}+z_{56}-2z_{67}-4z_{68})^{2}\\
 &  & +{\normalcolor {\color{red}\mathbf{2}}}(p_{3}+q_{3}+z_{57}+z_{67}-2z_{78}-4z_{79})^{2}+{\normalcolor {\color{red}\mathbf{2}}}(z_{68}+z_{78}+1-2z_{89})^{2}+{\normalcolor {\color{red}\mathbf{2}}}(z_{79}+z_{89}-1)^{2}.\nonumber 
\end{eqnarray}
}\end{widetext} whose properties are described in Table \ref{tab:factorizationTable511}.

\begin{center}
\begin{table}
\protect\caption{Energy landscape for a 17-qubit factorization of $511$ for various
Hamiltonians $H$ with equivalent ground states.\label{tab:factorizationTable511}}

\begin{centering}
\begin{tabular*}{1\columnwidth}{@{\extracolsep{\fill}}ccccccccccc}
\hline 
\noalign{\vskip2mm}
$H$ & $E_{{\rm gap}}=E_{|1\rangle}$ & $n_{1}$ & $E_{|2\rangle}$ & $n_{2}$ & $E_{|3\rangle}$ & $n_{3}$ & $E_{|4\rangle}$ & $n_{4}$ & $E_{|{\rm {max}\rangle}}$ & $\mathcal{R}$\tabularnewline[2mm]
\hline 
\noalign{\vskip2mm}
$H_{0}$ & $1$ & $2$ & $2$ & $20$ & $3$ & $60$ & $4$ & $113$ & $133$ & $17\,689$\tabularnewline
$H_{1}$ & $2$ & $2$ & $3$ & $8$ & $4$ & $4$ & $5$ & $16$ & $296$ & $10\,952$\tabularnewline
$H_{2}$ & $2$ & $2$ & $3$ & $12$ & $4$ & $8$ & $5$ & $35$ & $238$ & $7\,081$\tabularnewline[2mm]
\hline 
\end{tabular*}
\par\end{centering}

\centering{}\rule[-0.5ex]{1\columnwidth}{0.5pt}
\end{table}

\par\end{center}

\begin{center}

\par\end{center}

On this occasion we have managed to reduce the spectral ratio as well
as the number of local minima, as can be seen in the final column
of Table \ref{tab:factorizationTable511}.

\section{Conclusion}

We have presented two methods for manipulating the energy landscape
of a Hamiltonian without changing the ground state, and without adding
extra qubits: \emph{deducELM }and \emph{multiplicityELM. }We used
these methods to increase the size of the energy gap between the ground
state and lowest excited state of the Hamiltonian ($E_{{\rm gap}}$),
and to reduce the size of the gap between the ground state and highest
excited state ($E_{{\rm width}}$) for an adiabatic quantum computation
algorithm for integer factorization. 

For the first factorization example, deducELM was able to double $E_{{\rm gap}}$while
increasing $E_{{\rm width}}$ by only a factor of $\approx1.03$ (see
Table \ref{tab:factorizationTableDeduc}). We showed that for the
quantum computation to find a state that is a fixed distance away
from the state that encodes the solution to the problem, the runtime
is at most $\nicefrac{E_{{\rm width}}^{2}}{E_{{\rm gap}}^{3}}$, so
for this factorization example, the runtime would be decreased by
at most a factor of 7.54, or 754\% of the original maximum runtime.
For the second factorization example, we applied multiplicityELM and
were able to decrease the maximum runtime by at most 250\%.

Our method is in fact more general, in that it does not need to increase
$E_{{\rm gap}}$ or decrease $E_{{\rm width}}$. In general, it can
provide a new Hamiltonian whose ground state(s) are the same. One
may choose to manipulate the energy landscape of the Hamiltonian in
such a way to \emph{decrease} the size of $E_{{\rm gap}}$ or \emph{increase}
the size of $E_{{\rm width}}$, as would be useful for classical simulated
annealing algorithms such as the one described in \cite{Isakov2015}.
One may also choose to transform the Hamiltonian in such a way to
change the couplings between the qubits, so that it is easier to map
the Hamiltonian onto available hardware, such as in NMR-based AQC
where it may be difficult to find a molecule where the distances between
the spins match the couplings in the desired Hamiltonian.

\section*{Acknowledgments}

We would like to thank Emile Okada (Cambridge University, UK), Ilia
Zintchenko \& Matthias Troyer (ETH Zurich, Switzerland), and Ke Zj
\& Xinhua Peng (University of Science and Technology, China) for helpful
discussions.

\bibliographystyle{apsrev4-1}

\end{document}